\documentstyle[aps]{revtex}
\begin{document}
\draft
\title{$QCD_{1+1} $ in the Limit of a Large Number of Colors
and Flavors}
\author{Michael Engelhardt}
\address{Department of Condensed Matter Physics \\
Weizmann Institute of Science \\ Rehovot 76100, Israel }
\date{}
\maketitle

\begin{abstract}
QCD in 1+1 dimensions is examined in the limit of a large number of
colors and flavors. The Hamiltonian matrix is given in a Fock space
spanned by 't Hooft meson states and, for the case of zero fermion mass,
a submatrix is diagonalized numerically to give the low-lying
spectrum as a function of $N_F /N_C $. Pair creation effects generate
bound states which are complicated mixtures of components of different
meson number. There are a number of nontrivial zero modes; in the massive
part of the spectrum some states tend to a well-defined $N_F \gg N_C $
limit while others become unstable and disappear. The masses of most
states remain remarkably constant over a large range of $N_F /N_C $.
\end{abstract}

\pacs{   }

\section{Introduction}
Since its inception, the large $N_C $ expansion \cite{hoof1} \cite{hoof2}
has afforded a host of valuable insights into $QCD$ dynamics beyond
standard perturbation theory in the coupling constant. Besides providing
a solvable example of a nonabelian gauge theory in the case of one
space dimension \cite{hoof2}, it has been significant e.g. in clarifying
the relation between the Skyrme model and $QCD$ and in qualitatively
explaining the Zweig rule \cite{kleblec}. On the other hand, the number of
colors $N_C $ is not the only parameter of $QCD$ which can be regarded
as large. The same procedure may also be applied to the number of
flavors $N_F $. In the real world, especially at higher energies,
where the heavy quarks contribute to the dynamics, the effective
$N_F /N_C $ ratio becomes larger than unity. Thus it seems natural to
investigate in what way the presence of a large number of flavors
modifies the large $N_C $ picture and how those parts of large $N_C $
phenomenology which appear to be successes survive the modification.
Early work in this direction was done by Veneziano \cite{vene} and
some more recent explorations into the field of multiflavor gauge
theory have been carried out e.g. in connection with bosonization
treatments \cite{frish1} \cite{frishn}.

Particularly the low-dimensional models such as $QCD_{1+1} $ with a large
number of colors become considerably simplified in the framework of
light-cone quantization \cite{hoof2}. At this point it should
be remarked that the light-cone treatment corresponds in some respects to
an effective theory with well-defined limitations, as was elucidated
in \cite{lenz}. In the work presented here, no attention will be paid
to such questions. The results will be obtained entirely within the
usual naive light-cone framework and no firm claim is made as far as
their validity in normal coordinates is concerned. The prevailing
experience gives rise to the hope that the light-cone treatment does
yield correct excitation spectra.

The present foray into the multiflavor world focuses purely on the mesonic
spectrum of the model with zero quark mass matrix. The Hamiltonian is
calculated in a (truncated) basis of states made up of 't Hooft mesons,
which diagonalizes the Hamiltonian for $N_F =1$, $N_C \rightarrow \infty $.
Increasing the number of flavors such that $N_F /N_C $ remains finite
as $N_C \rightarrow \infty $ induces nonvanishing interactions between
the 't Hooft mesons; the evolution of the low-lying spectrum as a
function of $N_F /N_C $ can be followed by rediagonalizing the
Hamiltonian. Here, only the flavor singlet sector is discussed; however,
the generalization to finite flavor density presents no conceptual problem.
Since the numerical work up to now was carried out with limited
computational resources, only the one-meson and two-meson
(i.e. two- and four-parton) sectors could be included.

\section{The Mesonic Hamiltonian}
The Hamiltonian of multiflavor $QCD_{1+1} $ in the light-cone gauge with
zero quark mass matrix reads in momentum space
\begin{eqnarray}
H &=& -\frac{g^2 }{4\pi } \frac{N_C^2 -1}{N_C } \sum_{a,i}
\int \frac{dk}{k} : q_{ai}^{\dagger } (k) q_{ai} (k) : \nonumber \\
& & -\frac{g^2 }{8\pi } \sum_{a,b,i,j} \int \frac{dq}{q^2 } dk \,
dk^{\prime } \, \left( : q_{ai}^{\dagger } (k) q_{bi} (k^{\prime } +q)
q_{bj}^{\dagger } (k^{\prime } ) q_{aj} (k-q) : \right. \label{ham} \\
& & \ \ \ \ \ \ \ \ \ \ \ \ \ \ \ \ \ \ \ \ \ \ \ \ \ \
\left. + \frac{1}{N_C } : q_{ai}^{\dagger } (k) q_{ai} (k-q)
q_{bj}^{\dagger } (k^{\prime } ) q_{bj} (k^{\prime } +q) : \right)
\nonumber
\end{eqnarray}
Here, $i$ and $j$ are the color indices, $a$ and $b$ the flavor indices.
Note that the fermion fields do not carry a Dirac index, since one of
the components completely decouples in the light-cone gauge.
The Hamiltonian has already been normal-ordered, by applying Wick's
theorem, with respect to the
vacuum of the theory, which is the perturbative one:
\begin{eqnarray}
q_{ai}^{\dagger } (p) |0\rangle = 0 & \mbox{ for } & p<0 \\
q_{ai} (p) |0\rangle = 0 & \mbox{ for } & p>0
\end{eqnarray}
The singularities in the integrands finally are defined via the
principal value prescription. As already mentioned in the introduction,
all subleties associated with the singular nature of the light-cone
description are disregarded here; the reader is referred to \cite{lenz}.

The Hamiltonian (\ref{ham}) conserves baryon number, total momentum,
and overall color. Here, the focus will be on the color singlet
sector \cite{selbst} with zero baryon number and total momentum
denoted by $2K$. Due to the confining Coulomb potential, the physical
spectrum consists entirely of color singlet quark-antiquark bound
states, i.e. mesons, and conglomerates thereof \cite{barsg} \cite{calldg}.
It thus seems meaningful to consider the Hamiltonian matrix in a Fock
space spanned by color singlet mesonic excitations above the (perturbative)
vacuum. Mesons are created by the bilinear operators
\begin{equation}
M^{\dagger }_{Qnab} = \frac{1}{\sqrt{N_C } } \int_{0}^{Q} dp \,
f_n (Q,p) \sum_{i} q_{ai}^{\dagger } (p) q_{bi} (p-Q)
\end{equation}
where the $f_n $ are a complete set of bound state wave functions. A
particularly suitable such set are 't Hooft's wave functions,
\begin{equation}
f_n (Q,p) = \frac{1}{\sqrt{Q} } \phi_{n} (p/Q)
\end{equation}
where the $\phi_{n} $ satisfy the equation
\begin{equation}
-\left( \frac{1}{x} + \frac{1}{1-x} \right) \phi_{n} (x)
-\int \frac{dy}{(x-y)^2 } \phi_{n} (y) = \frac{\mu_{n}^{2} }{m_0^2 }
\phi_{n} (x)
\label{hoofeq}
\end{equation}
with the boundary condition
\begin{equation}
\phi_{n}^{\prime } (0) = \phi_{n}^{\prime } (1) = 0
\end{equation}
The mass scale $m_0 $ is given by
\begin{equation}
m_0^2 = \frac{g^2 }{2\pi } \frac{N_C^2 -1}{N_C }
\end{equation}
and $\mu_{n}^{2} $ is the invariant mass squared of the $n$-th
't Hooft meson. Apart from the exact zero-mode solution
$\phi_{0} (x) =1 $, a suitable orthonormal set in which to expand the
$\phi_{n} $ in order to solve (\ref{hoofeq}) numerically
is given in this case by cosine functions.

With this choice, the Hamiltonian is already diagonal in the basis defined
by the $M^{\dagger }_{Qnab} $ in the case that $N_F $ remains finite
and $N_C \rightarrow \infty $. Note that the operators $M^{\dagger } $
are not suitable for a rigorous bosonization of the theory since they
only obey canonical bosonic commutation relations up to additional terms
of the order $1/N_C $. It is now a straightforward, albeit lengthy,
procedure to calculate the Hamiltonian matrix by commuting through
quark operators. One obtains
\begin{eqnarray}
& & \langle 2K,n,a,b | H | 2K,m,c,d \rangle =
\delta_{ac} \delta_{bd} \delta_{nm} \frac{\mu_{n}^{2} }{4K} \label{me11} \\
& & \langle 2K,n,a,b | H | K+Q/2,m,c,d; K-Q/2,m^{\prime },c^{\prime },
d^{\prime } \rangle = \\
& & \ \ \ \ \ \ \ \ \ \ \ \ \ \ \ \
\frac{m_0^2 }{2} \frac{1}{\sqrt{N_C } } \frac{1}{(2K)^{3/2} }
\left[ \delta_{c^{\prime } d} \delta_{ac} \delta_{bd^{\prime } }
f_{mm^{\prime } n} \left( \frac{K+Q/2}{2K} \right)
+\delta_{cd^{\prime } } \delta_{ac^{\prime } } \delta_{bd}
f_{m^{\prime } mn} \left( \frac{K-Q/2}{2K} \right) \right] \nonumber \\
& & \langle K+Q/2,n,a,b; K-Q/2,n^{\prime },a^{\prime },b^{\prime } |
H | K+Q^{\prime } /2,m,c,d; K-Q^{\prime } /2,m^{\prime },c^{\prime },
d^{\prime } \rangle = \nonumber \\
& & \ \ \ \ \ \ \ \ \ \ \ \ \ \ \ \
\delta_{ac} \delta_{bd} \delta_{a^{\prime } c^{\prime } }
\delta_{b^{\prime } d^{\prime } } \delta_{nm}
\delta_{n^{\prime } m^{\prime } } \delta (Q-Q^{\prime } )
\left( \frac{\mu_{n}^{2} }{2K+Q} + \frac{\mu_{n^{\prime } }^{2} }{2K-Q}
\right)
+ \frac{m_0^2 }{8K^2 N_C } \times \label{me22} \\ & & \left[
\delta_{a^{\prime } c} \delta_{ac^{\prime } }
\delta_{bd} \delta_{b^{\prime } d^{\prime } }
f_{mm^{\prime } nn^{\prime } }^{EXC}
\left( \frac{K\! \! +\! Q/2}{K\! \! +\! Q^{\prime } /2} ,
\frac{K\! \! -\! Q/2}{K\! \! +\! Q^{\prime } /2} \right) +
\delta_{ac} \delta_{a^{\prime } c^{\prime } }
\delta_{b^{\prime } d} \delta_{bd^{\prime } }
f_{m^{\prime } mnn^{\prime } }^{EXC}
\left( \frac{K\! \! +\! Q/2}{K\! \! -\! Q^{\prime } /2} ,
\frac{K\! \! -\! Q/2}{K\! \! -\! Q^{\prime } /2} \right)
\right. \nonumber \\ & &
+ \delta_{a^{\prime } c^{\prime } } \delta_{ab^{\prime } }
\delta_{bd} \delta_{cd^{\prime } }
f_{mm^{\prime } nn^{\prime } }^{ANN}
\left( \frac{K\! \! +\! Q/2}{K\! \! +\! Q^{\prime } /2} ,
\frac{K\! \! -\! Q/2}{K\! \! +\! Q^{\prime } /2} \right) +
\delta_{ac^{\prime } } \delta_{a^{\prime } b}
\delta_{b^{\prime } d} \delta_{cd^{\prime } }
f_{mm^{\prime } n^{\prime } n}^{ANN}
\left( \frac{K\! \! -\! Q/2}{K\! \! +\! Q^{\prime } /2} ,
\frac{K\! \! +\! Q/2}{K\! \! +\! Q^{\prime } /2} \right) \nonumber \\ & &
+ \left. \delta_{a^{\prime } c} \delta_{ab^{\prime } }
\delta_{bd^{\prime } } \delta_{c^{\prime } d}
f_{m^{\prime } mnn^{\prime } }^{ANN}
\left( \frac{K\! \! +\! Q/2}{K\! \! -\! Q^{\prime } /2} ,
\frac{K\! \! -\! Q/2}{K\! \! -\! Q^{\prime } /2} \right) \! +
\delta_{ac} \delta_{a^{\prime } b}
\delta_{b^{\prime } d^{\prime } } \delta_{c^{\prime } d}
f_{m^{\prime } mn^{\prime } n}^{ANN}
\left( \frac{K\! \! -\! Q/2}{K\! \! -\! Q^{\prime } /2} ,
\frac{K\! \! +\! Q/2}{K\! \! -\! Q^{\prime } /2} \right) \right]
\nonumber
\end{eqnarray}
with the form factors (note that the $\phi_{n} $ are defined to be zero
outside the interval $[0,1] $)
\begin{eqnarray}
f_{mm^{\prime } n} (v) &=& \frac{1}{\sqrt{v(1-v)} } \int_{0}^{v} dx
\int_{0}^{1-v} dy \, \phi_{m} (x/v) \phi_{m^{\prime } } (y/(1-v) )
\frac{\phi_{n} (x) - \phi_{n} (v+y) }{(v+y-x)^2 } \nonumber \\
f_{mm^{\prime } nn^{\prime } }^{ANN} (v,w) &=&
\frac{(v+w)^2 }{\sqrt{vw(v\! +\! w\! -\! 1)} }
\int_{0}^{v} \! dx \int_{0}^{w} \! dy \,
\phi_{n} (x/v) \phi_{n^{\prime } } (y/w)
\frac{\phi_{m} (1\! +\! x\! -\! v)
\phi_{m^{\prime } } (y/(v\! +\! w\! -\! 1) ) }{(w+x-y)^2 } \nonumber \\
f_{mm^{\prime } nn^{\prime } }^{EXC} (v,w) &=&
\frac{(v+w)^2 }{\sqrt{vw(v+w-1)} } \int_{0}^{v} dx \int_{0}^{w} dy \,
\phi_{n} (x/v) \phi_{n^{\prime } } (y/w) \times \label{formf3} \\
& &
\frac{(\phi_{m} (1\! +\! x\! -\! v) -\phi_{m} (y) )
(\phi_{m^{\prime } } (x/(v\! +\! w\! -\! 1) ) - \phi_{m^{\prime } }
( (v\! +\! y\! -\! 1)/(v\! +\! w\! -\! 1) ) }{(v-1+y-x)^2 }
\nonumber
\end{eqnarray}
Here, in the kinetic part of the two-meson--two-meson matrix element,
a convention of definite ordering of the momenta in the states was
assumed, namely $Q,Q^{\prime } \ge 0 $. This avoids double-counting
two-meson states. Also, in this kinetic part, $1/N_C $-suppressed
terms in the normalization of the states were neglected. All interaction
parts, on the other hand, are valid for any $Q,Q^{\prime } $ and no
terms were neglected.

The processes described by $f_{mm^{\prime } n} $,
$f_{mm^{\prime } nn^{\prime } }^{ANN} $, and
$f_{mm^{\prime } nn^{\prime } }^{EXC} $ are associated with a flavor
flow represented diagrammatically in graphs (A),(B) and (C) of figure
(\ref{graphs}), respectively. These graphs clarify the large $N_F $
counting associated with each vertex; this will crucially influence
which contributions survive in the limit $N_F /N_C $ constant,
$N_C \rightarrow \infty $.

Note that the Hamiltonian can only change the number of mesons by one.
To change it by two, it would have to contain a contribution consisting
of four parton creation or annihilation operators with respect to the
vacuum; this, however, would completely commute through any meson
operators to the right or left, respectively, ultimately annihilating
the vacuum in relation to which the Hamiltonian is already normal-ordered.
Equations (\ref{me11})-(\ref{me22}) completely define the Hamiltonian
in the subspace containing up to two mesons. But for the fact that the
meson operators $M^{\dagger } $ only obey canonical commutation relations
up to terms of order $1/N_C $, the three-meson and four-meson interactions
described by (\ref{me11})-(\ref{me22}) would already represent the full
dynamics in mesonic language. This is because the Coulomb interaction,
as is clear from the original quark representation (\ref{ham}), can only
modify at most two out of an arbitrarily long string of meson operators
as it is being commuted through to eventually annihilate the vacuum.
However, the $1/N_C $-suppressed additional terms in the meson
commutators introduce additional interactions in larger clusters which
in general must be treated on the same footing as the vertices derived
above. Since the Hilbert space will here be truncated to the
one- and two-meson sectors, these terms need not be considered further
in the present calculation.

\section{The Flavor-Singlet Sector}
The above considerations still apply to an arbitrary structure in
flavor space; the results will now be specialized to the flavor singlet
sector. A flavor singlet meson can be constructed by the superposition
\begin{equation}
|Q,n\rangle_{S} = \frac{1}{\sqrt{N_F } } \sum_{a} M^{\dagger }_{Qnaa}
\end{equation}
whereas a flavor singlet two-meson state may consist either of two
singlet mesons or of two mesons carrying flavor spin one coupled to
a singlet:
\begin{eqnarray}
|Q,n;Q^{\prime }, n^{\prime } \rangle_{SS} &=&
\frac{1}{N_F } \sum_{a,b} M^{\dagger }_{Qnaa}
M^{\dagger }_{Q^{\prime } n^{\prime } bb} \\
|Q,n;Q^{\prime }, n^{\prime } \rangle_{TT} &=&
\frac{1}{N_F } \sum_{a,b} M^{\dagger }_{Qnab}
M^{\dagger }_{Q^{\prime } n^{\prime } ba}
\end{eqnarray}
Carrying out the flavor algebra, the Hamiltonian matrix in these states
takes the following form, where only terms which remain finite in the
limit $N_F /N_C $ finite, $N_C \rightarrow \infty $ are kept:
\begin{eqnarray}
& _{S}\langle 2K,n | H | 2K,m \rangle_{S} =
\delta_{nm} \frac{\mu_{n}^{2} }{4K} & \label{sing1} \\
& _{S}\langle 2K,n | H | K\! +\! \frac{Q}{2},m;
K\! -\! \frac{Q}{2},m^{\prime } \rangle_{SS} = 0 & \\
& _{S}\langle 2K,n | H | K\! +\! \frac{Q}{2},m; K\! -\! \frac{Q}{2},m^{\prime }
\rangle_{TT} =
\frac{m_0^2 }{2} \sqrt{\frac{N_F }{N_C } } \frac{1}{(2K)^{3/2} }
(1+(-1)^{m+m^{\prime } +n+1} )
f_{mm^{\prime } n} \left( \frac{K+Q/2}{2K} \right) & \\
& _{SS}\langle K\! +\! \frac{Q}{2},n; K\! -\! \frac{Q}{2},n^{\prime } | H |
K\! +\! \frac{Q^{\prime } }{2},m; K\! -\! \frac{Q^{\prime } }{2},m^{\prime }
\rangle_{SS} =
\delta_{nm} \delta_{n^{\prime } m^{\prime } } \delta (Q\! -\! Q^{\prime } )
\left( \frac{\mu_{n}^{2} }{2K+Q} \! +\!  \frac{\mu_{n^{\prime } }^{2} }{2K-Q}
\right) & \\
& _{TT}\langle K\! +\! \frac{Q}{2},n; K\! -\! \frac{Q}{2},n^{\prime } | H |
K\! +\! \frac{Q^{\prime } }{2},m; K\! -\! \frac{Q^{\prime } }{2},m^{\prime }
\rangle_{SS} =
\delta_{nm} \delta_{n^{\prime } m^{\prime } } \delta (Q\! -\! Q^{\prime } )
\left( \frac{\mu_{n}^{2} }{2K+Q} \! +\!  \frac{\mu_{n^{\prime } }^{2} }{2K-Q}
\right) & \\
& _{TT}\langle K\! +\! \frac{Q}{2},n; K\! -\! \frac{Q}{2},n^{\prime } | H |
K\! +\! \frac{Q^{\prime } }{2},m; K\! -\! \frac{Q^{\prime } }{2},m^{\prime }
\rangle_{TT} =
\delta_{nm} \delta_{n^{\prime } m^{\prime } } \delta (Q\! -\! Q^{\prime } )
\left( \frac{\mu_{n}^{2} }{2K+Q} \! +\!  \frac{\mu_{n^{\prime } }^{2} }{2K-Q}
\right) \nonumber \\
& + \frac{m_0^2 }{8K^2 } \frac{N_F }{N_C }
(1+(-1)^{m+m^{\prime } +n+n^{\prime } } )
\left[ f_{mm^{\prime } nn^{\prime } }^{ANN}
\left( \frac{K+Q/2}{K+Q^{\prime } /2} ,
\frac{K-Q/2}{K+Q^{\prime } /2} \right) +
f_{mm^{\prime } n^{\prime } n}^{ANN}
\left( \frac{K-Q/2}{K+Q^{\prime } /2} ,
\frac{K+Q/2}{K+Q^{\prime } /2} \right) \right] & \label{singn}
\end{eqnarray}
Here, the symmetry behavior of 't Hooft's wave functions $\phi_{n} (x) $
around $x=1/2$ was used, making explicit the selection rules for the
meson excitation quantum numbers. Physically, this is the manifestation
of CP invariance. These selection rules decouple states containing
single-meson components of even excitation number from those containing
single-meson components of odd excitation number. Note that this feature
does not persist in the flavor non-singlet sector.

The Hamiltonian matrix obtained above
exhibits in very transparent form the physics of the large $N_F $
world: Flavor singlets do not interact among each other. Nonvanishing
interactions only occur within flavor singlets, which resemble loops
built out of alternating color strings and flavor ``strings'' in the sense
that the two quarks at the end of a flavor string must be coupled to a
singlet. Within these loops, color strings may break into two, creating
a flavor string in between. Flavor strings are not necessarily binding;
however, the diagonalization of the Hamiltonian will show that there
are indeed bound states made up of more than one meson such that the
flavor string effectively acts as a bond. Note the remarkable
similarity of this picture to the model with one flavor, but adjoint
color \cite{kleb} \cite{kuta}.

In the original quark Hamiltonian, flavor does not appear as a quantum
number related to the interaction; thus, the notion of a flavor string
introduced above does not carry physical meaning in the same sense as
the color string, which in fact represents a region of nonvanishing
chromoelectric field energy density. The flavor string ``force''
is generated entirely by the quantum fluctuations,
which manifest themselves in the diverging polarizability of the vacuum
as $N_F \rightarrow \infty $. The weakness of the color forces between
the mesons is offset by the ease with which quark-antiquark
pairs are created. Formally, this is reflected in the fact that
only graphs which imply the existence of free flavor indices to be
summed over have a chance to compensate the usual $1/N_C $-suppression
of meson vertices by powers of $N_F $ in the numerator. These are
exactly the graphs in which a (flavor singlet) quark-antiquark pair
is created or annihilated, i.e. a color string is broken into two
or vice versa. Thus, only the graphs (A) and (B) in figure (\ref{graphs})
survive; only the form factors $f_{mm^{\prime } n} $ and
$f_{mm^{\prime } nn^{\prime } }^{ANN} $ enter the Hamiltonian matrix
(\ref{sing1})-(\ref{singn}).

In contrast to the one-flavor theory, where pair creation is not enhanced,
and which is simply solved by 't Hooft's mesons, the bound state spectrum
will now consist of superpositions of states containing different numbers
of mesons. In the case of the one-flavor, adjoint color theory \cite{kleb}
it was found that the low-lying spectrum is still very well described
by a truncation to few quark-antiquark pairs. This gave rise to the hope
that the restriction to one- and two-meson states imposed in the present
calculation would also give an adequate description of the first few
bound states. Ultimately, this point will have to be checked by enlarging
the Hilbert space.

Numerical evaluation and diagonalization of the
Hamiltonian matrix (\ref{sing1})-(\ref{singn}) gives the low-lying
spectrum as a function of $N_F /N_C $. In the actual calculation, the
two-meson states were supplied with a relative momentum wave function,
i.e. the physical basis used was of the form (cf. the convention
introduced further above of definite ordering of momenta in the kets):
\begin{equation}
\frac{1}{\sqrt{2K} } \int_{0}^{2K} \! \! dQ \,
\psi_{l} (Q/2K) \frac{1}{\sqrt{2} } \left(
| K\! +\! Q/2,m; K\! -\! Q/2,m^{\prime } \rangle_{TT} \pm
| K\! +\! Q/2,m^{\prime } ; K\! -\! Q/2,m \rangle_{TT} \right)
\end{equation}
where the $\psi_{l} $ were chosen to be (properly normalized) cosines
(with the plus sign) and sines (with the minus sign). Note that for
$m=m^{\prime } $ only even (cosine) relative momentum wave functions are
possible due to the Bose symmetry. Note also that the form factors
(\ref{formf3}) contain logarithmic divergences for
zero quark mass which pose no problem if a continuous relative momentum
wave function is introduced as above, whereas they make a direct
discretization of the momentum difficult. The basis used in the
computation consisted of eight one-meson and 26 two-meson states.
Bound states were identified by subtracting the expectation value
of the kinetic energy from the energy eigenvalue and checking for a
negative result.

\section{Numerical Results}
The numerical calculation yielded four nontrivial zero modes besides the
zeroth 't Hooft meson $|2K,0\rangle_{S} $, which is an eigenstate
of the Hamiltonian for all $N_F /N_C $, as can be easily seen from the
three-meson form factor $f_{mm^{\prime } n}$ (cf. (\ref{formf3})).
Furthermore, a spectrum of massive
bound states, plotted in figure (\ref{spec}) as a function of $N_F /N_C $,
was found. Some of the massive states disappear for intermediate values
of $N_F /N_C $ and reappear with very similar energy and mesonic
composition. In such cases the range where the state disappears is
bridged in figure (\ref{spec}) by the dotted lines to indicate the kinship
between the spectral trajectories in different regions.

The different classes of states obtained can each be described in more
detail as follows. The nontrivial zero modes are of the structure
\begin{equation}
\sum_{n odd} a_n | 2K,n \rangle_{S}
+\frac{1}{\sqrt{K } } \int_{0}^{2K} dQ \, \chi (Q/2K)
|K+Q/2,0; K-Q/2,0 \rangle_{TT}
\label{zmfo}
\end{equation}
where the restriction to odd $n$ comes simply from the selection rules
already discussed above in connection with equations
(\ref{sing1})-(\ref{singn}). It should be noted that the Hilbert space
used in the present calculation contained the first four relative
momentum states made up of two ground state 't Hooft mesons and also the
first four single-meson states of odd excitation number. In this basis,
four nontrivial zero modes were found. It thus seems natural to
expect that the full theory in fact contains an infinity of zero modes
of the form (\ref{zmfo}). For small $N_F /N_C $, the one-particle
admixture vanishes as $N_F /N_C $. Such a scaling behavior is to be
expected; for $N_F /N_C \rightarrow 0 $, the Hamiltonian becomes diagonal
in the mesonic basis and thus the eigenstates are dominated by a single
(either one- or two-particle) Fock state, with $N_F /N_C $-suppressed
admixtures obtainable in perturbation theory. The same behavior is
exhibited by the binding energies. This feature of course also arises
in the massive part of the spectrum to be discussed further below.

At $N_F =N_C $, the one-particle content ranges from 40\% for states
in which these one-particle admixtures are made up of low-lying
't Hooft mesons, downwards as more excited mesons participate. The binding
energies at $N_F /N_C $ are exactly $2m_0^2 /K $ (up to the numerical
accuracy of the calculation) for three of the zero modes, rising to
$2.5 m_0^2 /K $ for the fourth. In the opposite limit of large $N_F /N_C $
on the other hand, the two-meson admixture is suppressed as $N_C /N_F $
and the states become mixtures of odd single 't Hooft meson states.
As before, this scaling is easy to understand: For large $N_F /N_C $,
the two-meson--two-meson interaction Hamiltonian, which is positive
(this will be discussed further below), becomes dominant, and thus it is
not favorable for a state to contain a high two-meson proportion.
In order to maintain a meaningful balance between the different parts
of the Hamiltonian, the two-meson amplitude must be proportional to
$\sqrt{N_C /N_F } $, i.e. the two-meson content of the states must
scale as $N_C /N_F $. Again, this property of course persists for the
massive part of the spectrum. The binding energies of the zero modes
must accordingly approach constants. Numerically one obtains two states
with a binding energy of $8m_0^2 /K $ for large $N_F /N_C $, one with
binding energy $20m_0^2 /K $, and one with $29m_0^2 /K $. It has not
been possible to give analytical solutions for the nontrivial zero modes;
there must be subtle dynamical mechanism responsible for suppressing
transitions from the single-meson admixtures in (\ref{zmfo}) to the
massive part of the spectrum. It should be noted that some of the
massive bound states to be discussed below also contain sizeable
admixtures of odd single-meson states.

Continuing on to the massive part of the spectrum, the states containing
odd single-meson components appear as the third and the fourth massive
state in the $N_F /N_C \ll 1 $ region (cf. figure (\ref{spec})). In
detail, these two states however exhibit some differences: The fourth
state only exists at low $N_F /N_C $, and in the region shown consists of
about equal parts of $|5\rangle $ and $|1,1\rangle $ (the quantum
numbers in the kets merely indicate the 't Hooft meson content, whereas the
momentum quantum numbers have been suppressed). For very low $N_F /N_C $,
though, the state becomes dominated by the $|1,1\rangle $ part. By contrast,
the third massive state becomes dominantly a one-meson state in the
$N_F /N_C \rightarrow 0 $ limit, namely a $|3\rangle $. As $N_F /N_C $
rises, it acquires admixtures of $|0,0\rangle $, $|1,1\rangle $ and
$|0,2\rangle $ so that the $|3\rangle $-component is still found with
55\% probability at $N_F =N_C $. Whereas all other states found are
dominated by two to three Fock components even in the transition region
around $N_F =N_C $, this state becomes a strong mixture. This is
especially pronounced where the state becomes unstable. Remarkably,
as $N_F /N_C $ rises beyond unity, the two-meson content rises further
to reach 96\% at $N_F /N_C =50$; one third of this is the
$|0,0\rangle $ admixture. In accordance with the argument given further
above, this ultimately upsets the balance between potential and
kinetic energies. Thus, the state is very weakly bound above $N_F /N_C =10$
and finally disappears around $N_F /N_C =75$ instead of increasing its
one-particle content to give a well-defined $N_F /N_C \rightarrow \infty $
limit. The large range over which this state survives is remarkable,
though.

On the other hand, the rest of the massive states shown in figure
(\ref{spec}) contain even single-meson states. Here again, one finds
states which become dominated by a two-meson Fock component for
$N_F /N_C \rightarrow 0$ (namely the second massive state, which
becomes predominantly a $|0,1\rangle $) and others dominated by a
one-meson component in this limit: The first massive state becomes a
$|2\rangle $, the fifth an $|8\rangle $. The one-particle fraction
of the first massive state falls to about two thirds at $N_F =N_C $ due to
admixtures of $|0,1\rangle $ and then rises again to give a sensible
$N_F /N_C \rightarrow \infty $ limit. Remarkably, in this region, its mass
falls to zero. Apart from this last feature, the fifth massive state
behaves very similarly, acquiring a $|1,2\rangle $ admixture of about
one third in an intermediate regime of $N_F /N_C $ and then becoming
again a pure $|8\rangle $. The second massive state acquires a steadily
rising admixture of $|4\rangle $ (about 25\% at $N_F =N_C $) to
rapidly become pure $|4\rangle $ for large $N_F /N_C $. The
state which is predominantly a $|6\rangle $ for $N_F /N_C \rightarrow
\infty $ only appears at $N_F /N_C >10$; it is the fourth massive state
in this region (cf. figure (\ref{spec})). This behavior is most
probably due to the truncation of Hilbert space, i.e. the state may
be dominated at lower $N_F /N_C $ by a two-meson component not included
in the calculation.

Summing up, the gross features of the spectrum can be characterized
as follows: On the one hand, there are the zero modes, which consist
predominantly of $|0,0\rangle $ with growing admixtures of odd
single-meson states as $N_F /N_C $ rises until these admixtures
become dominant for large $N_F /N_C $. Concomitantly, the massive
states containing odd single-meson components become unstable as
$N_F /N_C $ is increased, which is signalled by these states becoming
strong mixtures without a dominant Fock component. On the other hand, the
massive states containing even single-meson admixtures display a
well-defined large $N_F /N_C $ limit, where they are dominated by just
these even single-meson states. The lowest of these states becomes
massless for $N_F /N_C \rightarrow \infty $. The reason for this
exceptional behavior is unclear; all other states display a remarkably
stable mass over a wide range of $N_F /N_C $. Also this stability of
the masses is a puzzle; the binding energies of the states on their own
vary quite strongly, as is evident from the values indicated in
figure (\ref{spec}).

\section{Comments and Outlook}
Some comments are in order regarding the results presented above. The
first of these concerns the low $N_F /N_C $ limit. The reader will
have noticed that not all of 't Hooft's mesons appear in the bound state
spectrum in this limit. To be precise, only $|0\rangle $, $|2\rangle $,
$|3\rangle $ and $|8\rangle $ are present; $|1\rangle $, $|4\rangle $,
$|5\rangle $, $|6\rangle $ and $|7\rangle $ are missing. These states
of course do appear in the numerical diagonalization, with small
admixtures of two-meson states, but with a positive expectation value of
the potential energy. Thus they are not bound and must rather be viewed
as resonances. For very small $N_F /N_C $, this distinction becomes
meaningless. An experimenter in 1+1 dimensions, working in a finite
laboratory for a finite time, will observe all of 't Hooft's mesons as
$N_F /N_C \rightarrow 0$. Strictly speaking, however, if for fixed
$N_F /N_C $, the experimenter waits long enough, he/she will only
find some of these mesons as stable particles. In the case of the
higher mesons, the picture may be strongly influenced by the truncation
of Hilbert space, but the absence of the $|1\rangle $ at least seems
conspicuous. On the other hand, the bound states become very weakly bound
for $N_F /N_C \rightarrow 0$, so that the experimenter would indeed observe
the $|0,1\rangle $ (this is the second massive state) as two essentially
independent ground state and first excited 't Hooft mesons, respectively,
in a finite experiment.

As a further comment, it should be stressed that, to generate
bound states in the model, it is crucial to combine
one-meson and two-meson states. Diagonalizing the two-meson sector
on its own does not give binding, i.e. the two-meson--two-meson
interaction matrix is positive, as mentioned above. This seems plausible,
for the following reason: If there was a pure two-meson state
with a negative expectation value of the interaction, then one would
have a state with arbitrarily negative overall energy by simply choosing
$N_F /N_C $ suitably large such that the interaction dominates the
kinetic part. On the other hand, the ground state (vacuum) energy
of the theory has been defined to be zero, so that a state with
negative energy should not be possible. Conversely, including
one-meson states can very well generate binding through the three-meson
vertex, and the energy is stabilized to remain positive by the
two-meson--two-meson vertex; this is indeed precisely what happens, as
mentioned above in the detailed discussion of the spectrum at large
$N_F /N_C $. If one writes a mesonic Hamiltonian in terms of the
operators $M^{\dagger } $ and their adjoints such that it correctly
reproduces the Hamiltonian matrix (\ref{me11})-(\ref{me22}),
it is the cubic term that may become negative, and the quartic term
which stabilizes the theory.

Summing up, the large $N_C $, large $N_F $ theory contains a considerably
wealthier phenomenology than the one-flavor theory. It represents a
more sophisticated model for mesons, incorporating pair creation
effects and thus mimicking the presence of a meson cloud, thought to
be an important ingredient of three-dimensional hadrons.
Despite the complex behavior of the model,
many of its features seem to be susceptible to numerical analysis,
given the power of modern-day computers. In this respect, a
significant improvement on the treatment presented here should be
possible, since the latter only made use of two workstations.
By using a mesonic basis, which already incorporates a sizeable
part of the dynamics, the bulk of the computation lies in the
evaluation of the Hamiltonian matrix rather than its diagonalization.
The matrix elements, however, can be evaluated completely independent
of each other, making a large improvement in speed possible by simply
using parallel processors. Including sectors with a higher number of
mesons demands the computation of higher-dimensional integrals; this
will only be feasible beyond a certain point by employing Monte Carlo
methods.

Besides improvements of the numerical aspects, there are a number
of other extensions of this work which seem to merit attention. It
would be interesting to understand in more detail the dynamics leading
to binding. This endeavor would profit greatly from some analytical
understanding of the nontrivial zero modes. Especially the zero mode
sector may to some extent be tractable analytically, at least in some
approximation, since the form factors simplify considerably
if some of the participating mesons are ground-state ones.

Furthermore, the approach followed here becomes the most questionable
in the high $N_F /N_C $ limit. There, the coupling between sectors
of different particle number is the strongest and the inclusion of
Fock components with three or more mesons could substantially alter
the results. It would be helpful to devise an approximation scheme
which is particularly suited for this potentially interesting new limit,
i.e. one which makes explicit use of the smallness of the parameter
$N_C /N_F $.

With a view to approaching more realistic theories, it may be helpful
to work out the perturbative corrections in $1/N_C $. Finally, note also
that the model is easily generalized to non-singlet flavor (flavor
spin one objects will resemble open strings) as well as
non-zero quark masses. In fact, in the latter case, the Hamiltonian
matrix retains exactly the form (\ref{me11})-(\ref{me22}); only the
't Hooft wave functions entering the form factors and the corresponding
meson masses are modified. It is to be expected that quark masses
suppress pair creation related effects. Thus, the physical spectrum
should remain closer to the 't Hooft one for given $N_F /N_C $.
The zero quark mass model considered in this work is the case
furthest removed from the well-studied one-flavor theory, since it
maximally allows for the pair creation processes which represent
the essence of the model's nontrivial dynamics.

\subsection*{Acknowledgements}
The author wishes to thank S.Levit and A.Schwimmer for motivating
discussions. This work was supported by a MINERVA fellowship.

\begin{figure}
\caption{Flavor flow associated with the form factors
$f_{mm^{\prime } n} $, $f_{mm^{\prime } nn^{\prime } }^{ANN} $ and
$f_{mm^{\prime } nn^{\prime } }^{EXC} $, respectively.}
\label{graphs}
\end{figure}

\begin{figure}
\caption{Invariant square masses of the massive bound states as a function
of $N_F /N_C $. Dotted lines are merely to guide the eye in connecting
spectral trajectories of states which disappear at intermediate
values of $N_F /N_C $. The figures indicate the binding energies of
the states at selected points in the same units, i.e.
$K\langle H_{pot} \rangle /2m_0^2 $.}
\label{spec}
\end{figure}


\begin{references}
\bibitem{hoof1} G.'t Hooft, Nucl. Phys. B72 (1974) 461
\bibitem{hoof2} G.'t Hooft, Nucl. Phys. B75 (1974) 461
\bibitem{kleblec} I.Klebanov, lectures presented at the workshop of the
Graduiertenkolleg Erlangen-Regensburg (Germany) on ``QCD and Hadron
Structure'', 9.6.-11.6.1992 at Kloster Banz; GK-Notes 3-92
\bibitem{vene} G.Veneziano, Nucl. Phys. B117 (1976) 519
\bibitem{frish1} D.Gepner, Nucl. Phys. B252 (1985) 481
\bibitem{frishn} G.D.Date, Y.Frishman and J.Sonnenschein, Nucl. Phys.
B283 (1987) 365
\bibitem{lenz} F.Lenz, M.Thies, K.Yazaki and S.Levit, Ann. Phys. (N.Y.)
208 (1991) 1
\bibitem{kleb} K.Demeterfi, I.Klebanov and G.Bhanot, Nucl. Phys. B418
(1994) 15; \\ G.Bhanot, K.Demeterfi and I.Klebanov, Phys. Rev. D48
(1993) 4980
\bibitem{kuta} D.Kutasov, Nucl. Phys. B414 (1994) 33
\bibitem{selbst} M.Engelhardt and B.Schreiber, ``Elimination of Color
in 1+1-dimensional $QCD$'', to appear in Z. Phys. A
\bibitem{barsg} I.Bars and M.B.Green, Phys. Rev. D17 (1978) 537
\bibitem{calldg} C.G.Callan, N.Coote and D.J.Gross, Phys. Rev. D13
(1976) 1649
\end{references}
\end{document}